# Communication Architecture for Autonomous Power-to-X Platforms: Enhancing Inspection and Operation With Legged Robots and 5G

Peter Frank
*Institute of Industrial Automation and Software Engineering*
*University of Stuttgart*
Stuttgart, Germany
peter.frank@ias.uni-stuttgart.de

Falk Dettinger
*Institute of Industrial Automation and Software Engineering*
*University of Stuttgart*
Stuttgart, Germany
falk.dettinger@ias.uni-stuttgart.de

Daniel Dittler
*Institute of Industrial Automation and Software Engineering*
*University of Stuttgart*
Stuttgart, Germany
daniel.dittler@ias.uni-stuttgart.de

Pascal Häbig
*Institute of Energy Economics and Rational Energy Use*
*University of Stuttgart*
Stuttgart, Germany
pascal.haebig@ier.uni-stuttgart.de

Nasser Jazdi
*Institute of Industrial Automation and Software Engineering*
*University of Stuttgart*
Stuttgart, Germany
nasser.jazdi@ias.uni-stuttgart.de

Kai Hufendiek
*Institute of Energy Economics and Rational Energy Use*
*University of Stuttgart*
Stuttgart, Germany
kai.hufendiek@ier.uni-stuttgart.de

Michael Weyrich
*Institute of Industrial Automation and Software Engineering*
*University of Stuttgart*
Stuttgart, Germany
michael.weyrich@ias.uni-stuttgart.de

*Abstract—* Inspection and maintenance of offshore platforms are associated with high costs, primarily due to the significant personnel requirements and challenging operational conditions. This paper first presents a classification of Power-to-X platforms. Building upon this foundation, a communication architecture is proposed to enable monitoring, control, and teleoperation for a Power-to-X platform. To reduce the demand for human labor, a robotic system is integrated to autonomously perform inspection and maintenance tasks. The implementation utilizes a quadruped robot. Remote monitoring, control, and teleoperation of the robot are analyzed within the context of a 5G standalone network. As part of the evaluation, aspects such as availability and latency are recorded, compared, and critically assessed.

## I. Introduction

The global challenges posed by climate change and the quest of transforming energy supply to pure renewable sources call for new, sustainable solutions to provide new molecular energy carriers based on renewables. Power-to-X (PtX) technologies, particularly those based on green hydrogen[1], are emerging as a cornerstone of future energy systems. Due to higher potentials and less landscape disturbance, offshore wind energy, presents a promising opportunity to produce climate-neutral electricity at scale and to convert it directly into hydrogen or PtX products such as ammonia, methanol, or other synthetic fuels at the point of generation, both as energy carriers of high energy density as well as chemical feedstock. [1]

Producing PtX products offshore, especially in off-grid environments, allows for the use of abundant wind resources while avoiding the high costs and energy losses associated with long-distance power transmission infrastructure [2]. Such systems are vital for decarbonizing energy-intensive sectors like in certain parts of chemical industry or in parts of the transport sector. As outlined by [3], hydrogen is not envisioned as an end-user fuel but rather as a central molecular energy carrier in a broader PtX economy that transforms renewable electricity into versatile products across sectors.

Moving PtX production offshore increases cost of such production compared to onshore, but at the same time reduces cost of power connection for offshore wind parks in case of a offgrid realization. The latter opens up as well to develop a vast range of sites worldwide without the possibility of a grid connection. However, such offshore offgrid PtX platforms are suggested to operate autonomously and therefore introduce a new technological paradigm. Their realization requires not only modular and scalable production units but also robust systems capable of reliable operation under extreme maritime conditions. As highlighted by [1], this involves significant challenges for system monitoring, fault detection, and maintenance, particularly in unmanned and remote configurations [1], [4].

Digital twins, advanced communication technologies, and autonomous robotic systems are considered key enablers in overcoming these challenges. Digital twins allow for real-time system representation and predictive maintenance, autonomous robots can take over inspection and repair tasks, and robust communication networks ensure safe coordination and data flow between subsystems [1], [5], [6].

---

[1] Green hydrogen means hydrogen produced by 100% renewable energy sources

The objective of this work is to develop a systematic understanding of the technological foundations required to enable such offshore production systems. Against this background, the central research question guiding this study is: *How can the safe and reliable operation of autonomous offshore PtX platforms be ensured?*

To contextualize this research question, Section II reviews the current landscape of PtX system implementations and technological concepts. In particular, recent demonstration projects highlight the diversity of system architectures currently being explored. Section III introduces an approach for implementing autonomous offshore PtX platforms, focusing on their realization and the communication technologies used. Finally, the results are evaluated in terms of communication performance and operational reliability.

## II. BACKGROUND AND RELATED WORK

The growing number of research and demonstration projects reflects a broad spectrum of architectures for offshore renewable-based PtX production systems. Projects such as FARWIND[7], Haru Oni[8], and H2Mare[9] explore diverse production system setups, from decentralized electrolyzers integrated into wind turbines to centralized, fully offshore platforms operating autonomously and off-grid.

This diversity in spatial distribution, grid connectivity, and degrees of automation illustrates the need for a systematic classification of PtX production systems. To provide a consistent analytical framework for evaluating communication and automation strategies, the following section introduces a classification of production system setups for green hydrogen and other PtX products using offshore wind power, structured by grid connection and production logic.

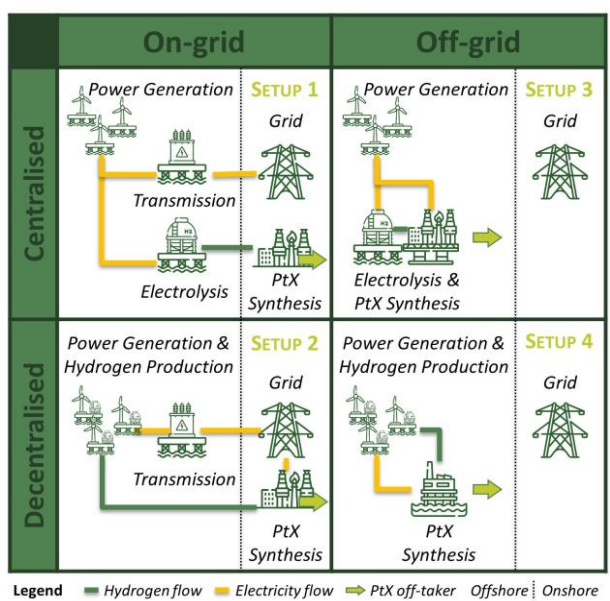

*Fig. 1 Classification of production system setups for green hydrogen using offshore wind power*

### A. Classification of Production System Setups for Green Hydrogen Using Offshore Wind Power

To systematically explore the technological requirements of such future production systems, Fig. 1 introduces a classification scheme of setups using two key dimensions: grid connectivity (on-grid vs. off-grid) and production logic (centralized vs. decentralized). This classification approach was initially introduced by [4], of which one of the authors is a contributor, and is expanded here with specific focus on communication and automation aspects, providing a graphical representation as a foundation for analyzing digital and autonomous capabilities.

In Setup 1, offshore electrolysis is centralized, allowing efficient direct wind energy utilization. The onshore synthesis remains grid-connected, simplifying communication demands but raising logistical challenges for hydrogen transport, notably requiring robust data exchange to monitor the offshore operation remotely.

A prominent conceptual study of this configuration is provided by the German research project *OffsH2ore*, which developed a modular, 500-Megawatt PEM electrolysis platform, designed for island-mode operation. [10]

Setup 2 decentralizes electrolysis by integrating compact units into individual wind turbines, necessitating extensive real-time communication for synchronization and monitoring. Projects such as H2Mare OffgridWind specifically investigate pipeline-based hydrogen transport, confronting complex control and data security challenges [9].

Setup 3 describes autonomous offshore PtX platforms with integrated electrolysis and synthesis units operating fully off-grid. The inherent isolation demands robust autonomous control systems and advanced real-time communication, highlighting critical challenges for such installations. Moreover, the uncrewed nature of these platforms renders conventional monitoring and maintenance infeasible.

The German H2Mare PtX-Wind hydrogen flagship project explores this setup, aiming to develop and demonstrate scalable concepts for decentralized offshore hydrogen production combined with centralized synthesis under real-world maritime conditions [9].

Setup 4 features decentralized electrolysis at individual turbines coupled with centralized offshore PtX synthesis, significantly amplifying complexity. The decentralized generation and centralized conversion require particularly advanced communication and cybersecurity solutions to ensure seamless operation.

Similar to Setup 3, this configuration is fully off-grid and suitable for deployment in areas with no supporting infrastructure. However, the complexity of coordinating multiple offshore generation nodes increases. Communication latency, synchronization of decentralized production, and cybersecurity concerns become more critical.

The presented classification framework provides a structured lens to evaluate how future PtX concepts using offshore wind can be realized. Building on [5], which focused on physical setups, this paper emphasizes digital and autonomous technologies. As these systems move towards high-autonomy, low-intervention operation, integrating Digital Twins, robust 5G communication, and autonomous robotics becomes essential, especially in remote environments beyond conventional maintenance and control capabilities.

It underlines as well that communication architecture differs substantially among the presented setups. Particularly in off-grid and centralized production setups, robust and secure real-time communication is paramount. The deployment of 5G networks, particularly 5G Standalone (5G SA), addresses these critical challenges, enabling seamless integration of digital twins, robotics, and decentralized control, thereby making fully autonomous offshore PtX platforms practically viable.

The production system design, as shown in the classification framework, serves mainly as an illustrative use case, while the focus is on the essential role of automation in realizing these platforms. The following section analyzes related work on these enabling technologies.[5, 6]

*B. Related Work*

Current research is heavily focused on the application of digital twins for PtX platforms [6], [11], [12]. These digital representations enable model-based control, covering various levels of hierarchy [4]. The primary focus is often on the implementation and modeling of physical or machine learning-based models to examine, the accuracy and computational efficiency of a methanation reactor model [13] or a hydrogen electrolysis system model [14]. These models contribute to process optimization by enabling more precise predictions and increasing the overall efficiency.

Beyond modeling, active data acquisition plays a central role in the concept of the digital twin [15]. Real-time operational data is essential for continuously improving models and adapting them to real-world conditions. In offshore environments, this presents a significant challenge, as data must be provided securely and in a timely manner.

A key aspect of secure data transmission is the network architecture. Established guidelines, such as those from the German Federal Office for Information Security (BSI) [16], the NIS Directive [17], and NAMUR [18] recommendations, provide valuable approaches for securing communication networks. These concepts include, among other things, a zoning approach that defines different security levels for network segments. By strictly separating production, monitoring, and external networks, the attack surface for cyber threats can be minimized.

Various communication technologies are available for transmitting data from offshore PtX platforms. Microwave links provide stable communication with nearby infrastructure points on land [19], while satellite communication serves as a backup or primary solution in remote areas [20]. Mobile communication technologies, such as 4G and increasingly 5G, offer an alternative, especially when offshore wind farms are equipped with corresponding private operated base stations [21]. The choice of the appropriate technology depends on factors such as latency, bandwidth, and availability.

To achieve efficient data and information exchange, a powerful communication system is required that provides the necessary bandwidth while also enabling time-critical applications. Currently, in addition to Long-Term Evolution (LTE), the communication technologies 5G Non-Standalone (5G NSA) and 5G SA are being used [22].

While LTE and 5G SA each utilize dedicated radio and backend components, 5G NSA combines the backend of LTE with the radio components of 5G SA [23]. This results in higher bandwidth and transmission speeds compared to classic LTE but does not take full advantage of the available capabilities of 5G SA [24].

Another crucial element is communication protocols for integrating collected data into existing systems. In the industrial context, OPC UA [25] has become a standard for interoperable and secure machine-to-machine communication. These protocols enable vendor-independent connectivity of sensors, control systems, and digital twins, ensuring seamless data processing and analysis.

Finally, the use of robotic assistance systems, such as robotic dogs, unmanned aerial vehicles (UVAs) and autonomous underwater vessels (AUVs), is being explored in offshore environments. These systems are equipped with sensors that enable autonomous data collection and inspection. They can detect leaks, visually monitor plant components, or take measurements in hard-to-reach areas. As a result, they contribute to improved monitoring and maintenance of PtX platforms while reducing the exposure of human workers to hazardous conditions. [5], [26], [27]

*C. Discussion and Conclusion*

While existing research extensively explores digital twins for PtX platforms, current studies primarily focus on modeling and simulation aspects, such as improving the accuracy and computational efficiency of physical or machine learning-based models. However, no known work presents a holistic concept for the secure and reliable operation of offshore PtX platforms by integrating digital twin technologies with active data acquisition and transmission. Ensuring safe and efficient operations in such complex environments requires not only precise modeling but also robust data management, secure communication, and adaptive control mechanisms—elements that remain fragmented in the current state of research.

Additionally, emerging technologies such as autonomous robots and 5G communication offer new opportunities for data acquisition and remote operation. Humanoid robots and robotic assistance systems could enhance inspection and maintenance processes through autonomous sensing and interaction capabilities [5]. Similarly, 5G networks promise ultra-low latency and high reliability [28], which are critical for real-time monitoring and remote intervention in offshore environments [29]. Despite these potentials, there is a lack of studies investigating latency, availability, and the practical feasibility of these technologies in offshore PtX applications.

The next section aims to address these gaps by presenting an integrated approach that combines digital twins, secure data acquisition, and emerging communication technologies to enable the autonomous and reliable operation of offshore PtX platforms.

III. APPROACH

This section first introduces the specific requirements in the context of offshore platforms and then presents the proposed approach.

## A. Requirements for the autonomous offshore plattform

The operation of self-sufficient offshore systems, as described in Section II, presents operators with complex challenges. These systems are often located far from shore and are left to their own devices. At present, the reliable operation of these systems is closely linked to the physical presence of specialists who are responsible for on-site inspections, maintenance work and the diagnosis and rectification of faults [26]. This leads to high logistical costs, safety risks and limited responsiveness in the event of a malfunction. With the increasing degree of automation, the use of autonomous systems will become necessary in the future, which will take over inspection and maintenance tasks and continuously record and evaluate the condition of the plant [5], [26], [27]. In this context, considerable amounts of operating data are generated - for example in the form of sensor data and visual information - which must be processed and made available in a suitable manner to support remote monitoring and control [27]. In addition, these systems often have to operate under highly fluctuating operating conditions, which are characterized by external influences such as weather and variable energy profiles [1]. This dynamic requires operations to be adapted flexibly and in a targeted manner in order to ensure both the efficiency and availability of the system. A key problem area here is the lack of a continuous connection to land-based infrastructures, which makes autonomous decision-making capability and high-performance wireless communication essential. The system status must be recorded locally and in real time, processed and made accessible to external users [13]. This enables well-founded operating decisions and process optimization based on current data. The following requirements for the concept of this article are derived from this:

- Active data collection and processing: The system must be able to continuously collect operational and environmental data and make it available for operational and diagnostic purposes.
- Autonomous operation and maintenance capability: To reduce the physical presence of personnel on site, the system must support autonomous systems for inspection and maintenance.
- Flexible communication infrastructure: Due to the isolated location, a flexible wireless communication solution must be implemented that covers both local and external requirements.
- Dynamic and adaptive operational management: Operation must be adaptable to changing external conditions in order to ensure continuous availability and efficiency.

## B. Proposed Approach

To address the requirements described above, a multi-layered concept is proposed that systematically integrates the necessary technical components and infrastructures (see Fig. 2).

An *Aggregation Server* is installed on the platform to ensure continuous and active data acquisition. This serves as a central collection point for all process-relevant data that originates from the *Process Control System* as well as from additional external sources, such as sensors on subsystems or measured values recorded by mobile robots. The aggregation server processes and stores this data locally and provides it for further applications. This creates a basis for supporting analytical processes such as status forecasts, diagnostic algorithms or anomaly detection. To reduce the physical presence of personnel on the platform, an autonomous robotic system is integrated into the overall concept. This system performs inspection and maintenance tasks and must be equipped with various sensors. The robot enables flexible data collection directly on the offshore components and transmits the information obtained to the aggregation server. This ensures comprehensive and continuous monitoring of the condition of the system.

Due to the lack of a physical network connection to the platform, wireless communication plays a key role. A local wireless infrastructure based on 5G technology is proposed within the platform and in the immediate vicinity. This enables the flexible connection of mobile units such as the autonomous robot as well as the real-time transmission of sensor data to the aggregation server and other local systems.

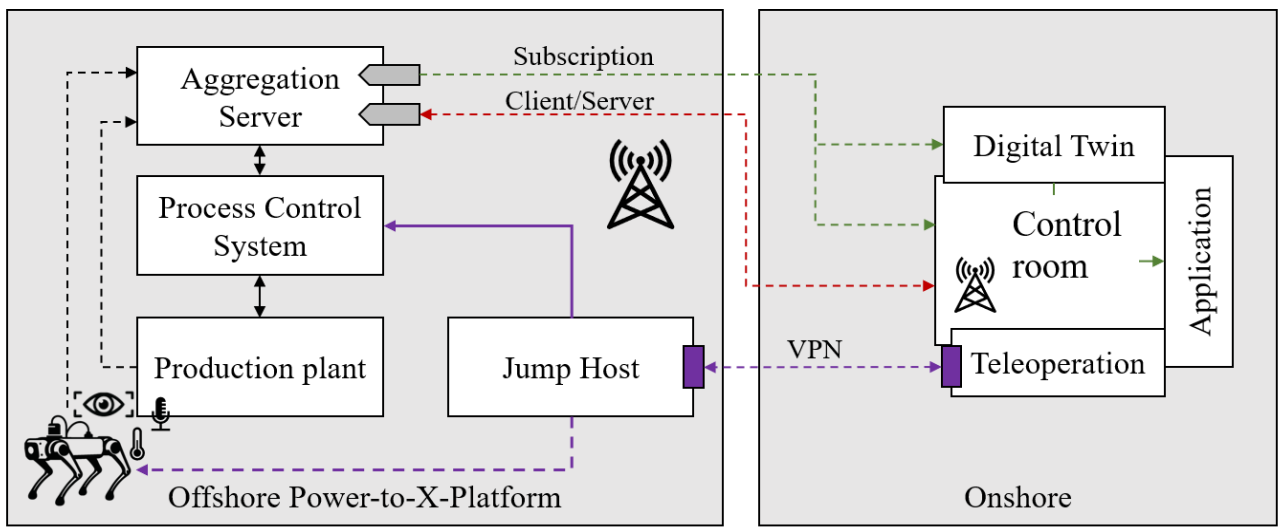

*Fig. 2. Communication architecture for offshore PtX platforms*

A multi-stage approach is recommended for the transmission of data to onshore-based systems: Both radio relay links and satellite-based communication solutions can be used for long-distance connections. The transmitted data flows into a central *Digital Twin*, which forms the basis for higher-level applications such as operational optimization based on weather and wind forecasts.

The distributed processing of data should be carried out using a combination of edge and cloud computing approaches. Edge devices can be used on the platform itself, which enable local and low-latency data processing, for example for the active analysis of robot data or process monitoring via the aggregation server. The advantages of edge computing lie particular in the rapid availability of analysis results and the reduced dependence on external communication. In addition, central cloud systems on land can be used, which enable long-term storage of large amounts of data, computationally intensive optimization algorithms and comprehensive analysis of different platforms. While edge systems are particularly impressive in time-critical scenarios, the cloud offers a higher degree of scalability and enables access to complex AI-supported applications.

For critical operations that need to be carried out remotely, a *Jump Host* must be established as a secure interface between the offshore platform and the onshore-based *Control Room*. The jump host serves as an access control point and ensures secure tunneling of all teleoperation processes. By using the jump host, sensitive control commands can be encrypted and transmitted securely, ensuring secure remote access to the platform. Particularly in conjunction with the robotic unit, the jump host enables manual interventions to be carried out securely when autonomous functions reach their limits or an immediate human decision is required in emergencies.

## IV. REALIZATION AND EXPERIMENTS

This chapter provides a brief introduction to the application scenario and discusses its implementation using a robot dog. Finally, the 5G experiments are presented

### A. Application Scenario

As outlined in Chapter 3, the developed robotic system is intended to play a central role in the automated execution of inspection and maintenance tasks on autonomous offshore platforms. A similar approach has been successfully field tested in an offshore setting [27]. The robot is designed to carry out routine inspection tasks such as monitoring fill levels and detecting wear or corrosion caused by the harsh maritime conditions. To detect thermal or acoustic anomalies like pipe temperatures indicating a malfunction or gas leaks the robot should be equipped with thermal cameras and microphones combined with AI models for anomaly detection. Routine inspection routes can be set through waypoints. For specific tasks an on-shore employee can control the robot via teleoperation. During these operations, sensor data is continuously aggregated and transmitted in accordance with the communication architecture described in Chapter 3. This opens up savings potential, by reducing the need for specialists on the platform to manually complete these tasks.

### B. Realization with a robotic dog

For the realization of the described application scenario, the Unitree Go1 EDU robot dog from the Institute of Industrial Automation and Software engineering at University of Stuttgart is used. This robot dog (see Fig. 3) is a relatively low-cost representative with opensource interfaces. With its compact dimensions of 645 x 280 x 400 mm, it is capable of operating in confined and complex environments [30]. It can traverse obstacles and stairs up to 12cm in height using a special stair climbing gait. Its onboard battery enables operating times of up to 2.5 hours [31].

In the 'EDU' version, open interfaces enable the development of customized software packages for special applications. It can carry up to 10 kg of additional load, which can be used for additional sensors and actuators such as an infrared camera or a robotic manipulator for inspection and maintenance tasks [30], [31]. These can be integrated via various hardware interfaces. Sufficient computing power for complex data processing is provided by three NVIDIA Jetson Nanos [30], [32]. The GO1 is therefore a cost-effective, flexible and expandable platform for analyzing autonomous inspection and monitoring use-cases [30], [31].

The robot dog is equipped with several sensors that enable autonomous navigation and allow it to carry out inspection tasks. Its five onboard wide-angle stereo cameras can be used for visual inspection. Additional sensors such as an infrared camera or microphone can be added to obtain further information for the inspection. The advantage of these additional sensors was demonstrated in [27].

In addition to the five stereo cameras, the robot can be equipped with a 2D LiDAR with integrated localization and obstacle avoidance. The wide-angle stereo cameras, positioned on the head, chin, on both sides and the underside, provide RGB images and depth information in the form of point clouds. (see Fig. 3) This enables comprehensive coverage of the robot's surroundings including the surface it is currently traversing. An integrated Inertial Measurement Unit (IMU) and foot pressure sensors further support terrain assessment. Special locomotion modes (walking, running, stairs climbing) allow adoption to different operational conditions and these can be optimized further via the open programming interfaces. [30], [32]

The Go1 can be controlled in various ways. It can autonomously navigate along predefined inspection routes or be manually teleoperated via a smartphone app or a remote control [30]. For teleoperation the robot can be controlled via a High-Level SDK, which allows to send high-level control commands like gait selection or the direction of movement [32]. For data transmission, the Go1 has a Wi-Fi interface and a 4G/5G module [30]. This allows the robot dog to carry out routine inspection rounds autonomously and be supported by a remote operator for special inspection tasks or complex situations. As mentioned in Section II, data transmission with 5G is analyzed below in order to address the research gap.

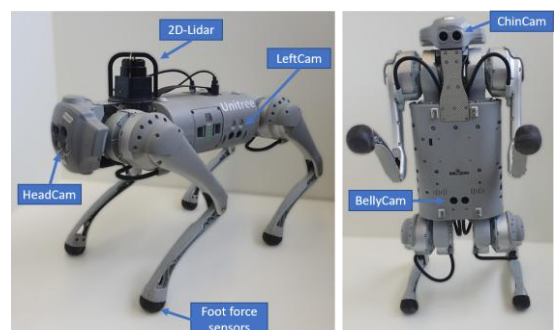

*Fig. 3. Unitree Go1 with sensors*

*C. 5G Experiments*

The following section examines whether deploying a function as a 'conventional function', 'containerized', or 'container-orchestrated' impacts processing and transmission time. Additionally, the performance of the different network types, LTE, 5G NSA, and 5G SA, is analyzed. The experiments regarding the LTE and 5G NSA networks are conducted in a Wide Area Network (WAN) provided by Deutsche Telekom, with a cloud in Mannheim serving as the backend server. In contrast, the experiments with 5G SA are conducted in a campus network provided by Nokia. The backend server in this case is represented by a laptop.

In both evaluation scenarios, a 5G M2 EVB Kit with RM520N-GL [33] manufactured by Quectel is used. The following parameters are used as evaluation metrics:

- **Latency:** Latency describes the transmission delay [34]. In this specific case, the latency encompasses both upload and download latencies.
- **Processing Time:** The processing time indicates the time required to process the input data and provide the results as variable [35]. It is directly affected by the hardware performance.
- **Packet Error Rate:** The PER indicates the number of incorrectly or not received packets [36]. Depending on the transmission protocol, a high PER prolongs the transmission duration or reduces the quality of the result, because packets need to be retransmitted or unavailable due to packet loss.
- **Packet Delivery Ratio:** The PDR indicates the number of correctly received packets [37].

Fig. 4 summarizes the transmission and processing times in milliseconds for the deployment of an algorithm on the cloud platform, represented as a box-and-whisker plot. The extensions '_docker', '_function', and '_kubernetes' indicate the deployment types, while '5G_nsa' refers to the communication type. All tests are performed using a cloud instance located in Mannheim with 60 GB of disk space, two virtual Central Processing Units, and 4 GB of Random Access Memory (RAM). The algorithm is executed 1,760 times per run, using images with an average size of 222.8 kB.

The measurements show that the mean processing times (Fig. 4 right) range between 244 ms and 248 ms, remaining comparable across deployment types. The transfer time has a median of approximately 310 ms for container-based deployments. By contrast, the median transmission times (Fig. 4 left) are about 230 ms for deployments as conventional functions and roughly 240 ms for deployments using container orchestration. Since the computer cluster also deploys functions within containers, it can be concluded that the variation in transmission times is due to fluctuations in network availability, with the deployment type having no significant impact on transmission and processing times for a service. However, the use of a computer cluster is recommended for scalability purposes.

The direct comparison of LTE, 5G NSA, and 5G SA is shown in Fig. 5. It should be noted that the measurements in the LTE and 5G NSA networks refer to WAN networks, while the expected latency in a campus network is actually lower. It is evident that the transmission latency in the LTE network, with a file size of 222.8 kB, is 150 ms in median. In contrast, the transmission latency for 5G NSA is 240 ms in median, and for 5G SA, approximately 70 ms. The superiority of LTE in the WAN domain is surprising and, in this case, can be attributed to the high dependency of performance on network load or sub-optimal network configuration of 5G NSA. It is evident that 5G SA is more performant, even in non-optimal scenarios as shown in our campus network, where no direct edge core access is available. Based on the measurements obtained, it can be concluded that, due to future viability, 5G SA communication should be taken into account for offshore-PtX platforms. The integrated features of extended Mobile Broadband (eMBB) communication and ultra Reliable and Low Latency Communication (uRLLC) enable the development of new use cases, such as, in our case, data exchange between autonomous units and the on-site computing hardware or the remote control of autonomous robots. Furthermore, 5G SA offers improved scalability, especially when it comes to the integration of Internet of Things (IoT) devices.

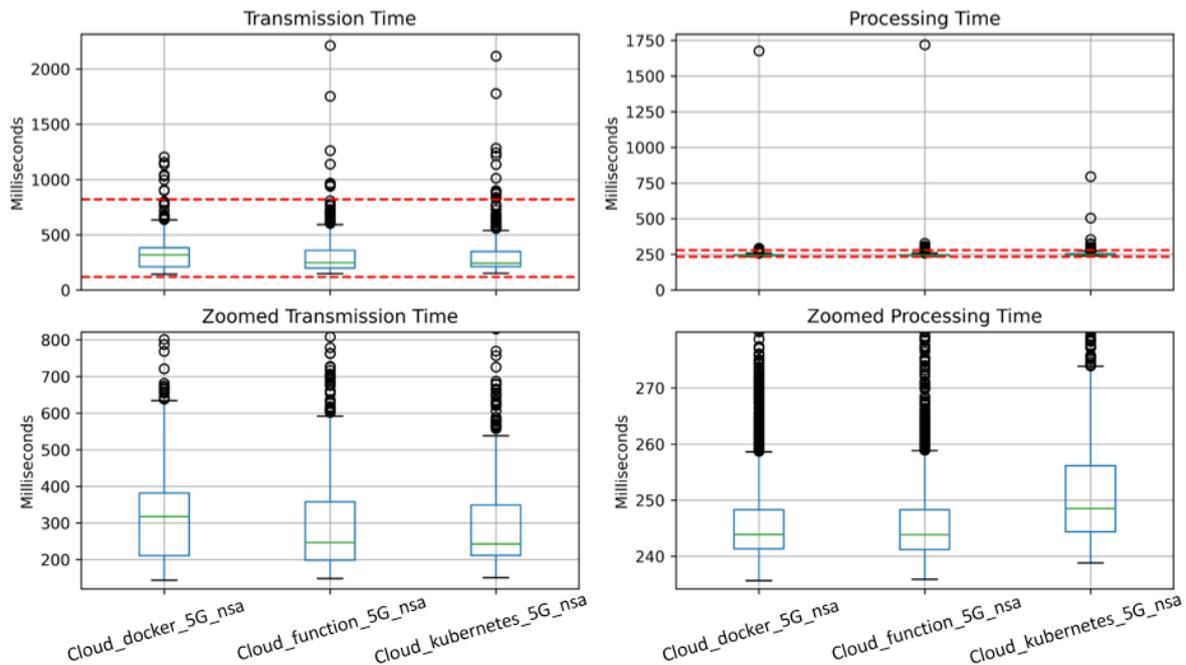

*Fig. 4. Investigation of the impact of the deployment strategy 'conventional function,' 'containerized,' and 'container-orchestrated' on transmission latency and the processing time of a function executed on a Cloud instance with two virtual CPU cores and 4GB of RAM.*

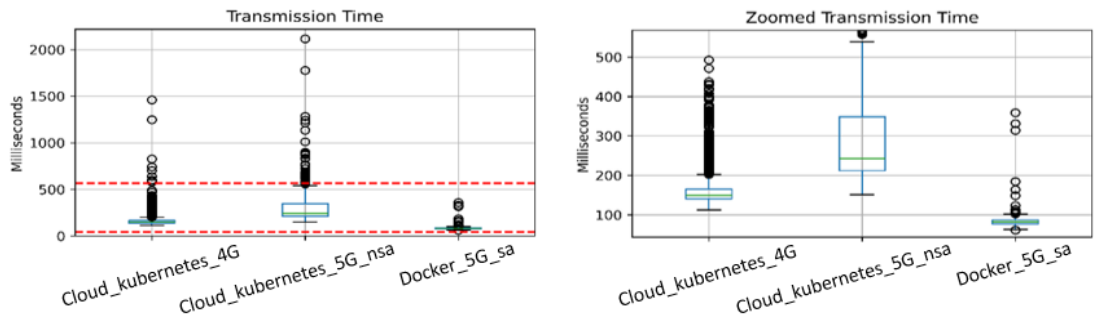

*Fig. 5. Comparison of the transmission latency of files with an average size of 222.8 kB in the Long Term Evolution (LTE) network (left), 5G non-standalone (middle), and 5G standalone (right). The LTE and 5G NSA networks are Wide Area Networks, whereas the 5G SA network is a campus network.*

## V. Discussion

The conducted experiments demonstrate that the Unitree Go1 robot serves as a suitable platform for evaluating data transmission and application scenarios in the context of inspection and maintenance tasks.

However, the deployment of such robotic systems in offshore Power-to-X environments requires further development, particularly with regard to robustness, sensor integration, and environmental resilience. One key aspect is the need for hardware enhancements to meet the demanding conditions of offshore environments. This includes compliance with safety and durability standards, such as IP67 certification, to ensure operational reliability under exposure to heavy rain, salt, and mechanical stress. For use on an industrial platform, industrial grade robodogs such as the ANYmal offer higher performance and resistance to harsh environmental conditions [27].

In addition, the integration of aerial drones could significantly extend the systems capabilities, enabling multi-perspective inspections and access to hard-to-reach areas. Furthermore, the current sensor configuration of the Go1 is limited in its ability to monitor critical process parameters. The integration of thermal imaging sensors is proposed to enable temperature measurements within pipelines and process equipment, which is particularly relevant for PtX applications involving thermal processes and fluid transport. Furthermore, the addition of manipulators such as robot arms could enable the robot to interact with its environment to actuate levers or switches and perform maintenance.

From a communication perspective, the use of a 5G standalone network resulted in an average teleoperation latency of approximately 100ms depending on the configuration and local network conditions. While this delay is within acceptable limits for remote control tasks, additional latency introduced by potential microwave (line-of-sight) communication links, necessary in off-grid offshore scenarios, must be considered. These aspects warrant further investigation, especially under realistic field conditions where network performance may vary.

It is important to note that all data were collected in an onshore setting. Although the scenarios were designed to be transferable, future work should validate the approach in actual offshore environments. In particular, the interaction between 5G networks and directional radio technologies needs to be examined in more detail to ensure seamless and reliable system integration.

Overall, the results indicate that mobile, autonomous robotic systems, supported by advanced communication architectures, hold significant potential for reducing manual labor and increasing operational safety in offshore PtX installations. However, targeted technological adaption and further validation efforts are essential for deployment.

## VI. Conclusion and Outlook

Especially in offshore settings, manual maintenance and inspection activities result in high personnel costs. This paper therefore addresses the automation of these tasks and presents the following results:

- Four classifications of production system setups for Green Hydrogen Using Offshore Wind Power
- A monitoring, control and teleoperation architecture for autonomous off-shore PtX plattforms.
- A realization with a robotic dog for inspection and maintenance using cameras and lidar sensors
- 5G standalone is well-suited for use cases on PtX platforms due to its high performance, the approaches for uRLLC and eMBB, and its expandability.

Future work will investigate the adaptation of autonomous robots to challenging environments. The focus will be on suitably combining or extending existing capabilities to handle unforeseen changes effectively.

## Acknowledgment

This contribution was funded by the Federal Ministry of Research, Technology and Space (BMFTR) under grant 03HY302R.

## References

[1] S. Kumar, T. Baalisampang, E. Arzaghi, V. Garaniya, R. Abbassi, and F. Salehi, "Synergy of green hydrogen sector with offshore industries: Opportunities and challenges for a safe and sustainable hydrogen economy," *J. Clean. Prod.*, vol. 384, p. 135545, Jan. 2023, doi: 10.1016/j.jclepro.2022.135545.

[2] B. Rego De Vasconcelos and J.-M. Lavoie, "Recent Advances in Power-to-X Technology for the Production of Fuels and Chemicals," *Front. Chem.*, vol. 7, p. 392, Jun. 2019, doi: 10.3389/fchem.2019.00392.

[3] C. Breyer, G. Lopez, D. Bogdanov, and P. Laaksonen, "The role of electricity-based hydrogen in the emerging power-to-X economy," *Int. J. Hydrog. Energy*, vol. 49, pp. 351–359, Jan. 2024, doi: 10.1016/j.ijhydene.2023.08.170.

[4] P. Häbig *et al.*, "A Modular System Architecture for an Offshore Off-grid Platform for Climate-neutral Power-to-X Production in H2Mare," *Procedia CIRP*, vol. 126, pp. 909–914, Jan. 2024, doi: 10.1016/j.procir.2024.08.348.

[5] D. Mitchell *et al.*, "A review: Challenges and opportunities for artificial intelligence and robotics in the offshore wind sector," *Energy AI*, vol. 8, p. 100146, May 2022, doi: 10.1016/j.egyai.2022.100146.

[6] D. Dittler *et al.*, "Digitaler Zwilling für eine modulare Offshore-Plattform: Effizienzsteigerung grüner Power-to-X-Produktionsprozesse," *Atp Mag.*, vol. 63, no. 6–7, pp. 72–80, Jun. 2022, doi: 10.17560/atp.v63i6-7.2606.

[7] A. Babarit, E. Body, J.-C. Gilloteaux, and J.-F. Hétet, "Energy and economic performance of the FARWIND energy system for sustainable fuel production from the far-offshore wind energy resource," in *2019 Fourteenth International Conference on Ecological Vehicles and Renewable Energies (EVER)*, May 2019, pp. 1–10. doi: 10.1109/EVER.2019.8813563.

[8] B. C. Leiss, "Chile's Energy Transition and the Role of Green Hydrogen".

[9] "Wasserstoff-Leitprojekte: H2Mare: Offshore-Technologien." Accessed: Apr. 15, 2025. [Online]. Available: https://www.wasserstoff-leitprojekte.de/leitprojekte/h2mare

[10] A. Rudolph *et al.*, "Projekt OffsH2ore Offshore-Wasserstofferzeugung mittels Offshore-Windenergie als Insellösung Endbericht," Fraunhofer-Institut für Solare Energiesysteme ISE, PNE AG, SILICA Verfahrenstechnik GmbH, KONGSTEIN GmbH, Wystrach GmbH 2023. Accessed: Apr. 16, 2025. [Online]. Available: https://www.ise.fraunhofer.de/content/dam/ise/de/documents/presseinformationen/2023/Projekt-OffsH2ore-Abschlussbericht-2023_public.pdf

[11] F. Bodenstein, S. Dieckmann, D. Dittler, A. Geschke, N. Jazdi, and M. Weyrich, "Digital Twin to Enhance Offshore Power-to-X Platforms with Operational Alarm Management," *Procedia CIRP*, vol. 130, pp. 780–785, Jan. 2024, doi: 10.1016/j.procir.2024.10.164.

[12] D. Dittler, D. Stauss, P. Rentschler, J. Stümpfle, N. Jazdi, and M. Weyrich, "Flexible Co-Simulation Approach for Model Adaption in Digital Twins of Power-to-X Platforms," in *2024 IEEE 29th International Conference on Emerging Technologies and Factory Automation (ETFA)*, Sep. 2024, pp. 1–4. doi: 10.1109/ETFA61755.2024.10710681.

[13] L. Peterson, J. Bremer, and K. Sundmacher, "Challenges in data-based reactor modeling: A critical analysis of purely data-driven and hybrid models for a CSTR case study," *Comput. Chem. Eng.*, vol. 184, p. 108643, May 2024, doi: 10.1016/j.compchemeng.2024.108643.

[14] P. Rentschler, C. Klahn, and R. Dittmeyer, "The Need for Dynamic Process Simulation: A Review of Offshore Power-to-X Systems," *Chem. Ing. Tech.*, vol. 96, no. 1–2, pp. 114–125, 2024, doi: 10.1002/cite.202300156.

[15] G. Hildebrandt, D. Dittler, P. Habiger, R. Drath, and M. Weyrich, "Data Integration for Digital Twins in Industrial Automation: A Systematic Literature Review," *IEEE Access*, vol. 12, pp. 139129–139153, 2024, doi: 10.1109/ACCESS.2024.3465632.

[16] "Bundesamt für Sicherheit in der Informationstechnik," Bundesamt für Sicherheit in der Informationstechnik. Accessed: Apr. 16, 2025. [Online]. Available: https://www.bsi.bund.de/DE/Home/home_node.html

[17] "Umsetzung der NIS-2-Richtlinie für die regulierte Wirtschaft," Bundesamt für Sicherheit in der Informationstechnik. Accessed: Apr. 15, 2025. [Online]. Available: https://www.bsi.bund.de/DE/Themen/Regulierte-Wirtschaft/NIS-2-regulierte-Unternehmen/nis-2-regulierte-unternehmen.html?nn=1115626

[18] "NAMUR – Interessengemeinschaft Automatisierungstechnik der Prozessindustrie e.V." Accessed: Apr. 15, 2025. [Online]. Available: https://www.namur.net/de/

[19] L. I. Gliga, R. Khemmar, H. Chafouk, and D. Popescu, "A Survey of Wireless Communication Technologies for an IoT-connected Wind Farm," *Wirel. Pers. Commun.*, vol. 122, no. 3, pp. 2253–2272, Feb. 2022, doi: 10.1007/s11277-021-08991-2.

[20] M. A. Ullah, K. Mikhaylov, and H. Alves, "Enabling mMTC in Remote Areas: LoRaWAN and LEO Satellite Integration for Offshore Wind Farm Monitoring," *IEEE Trans. Ind. Inform.*, vol. 18, no. 6, pp. 3744–3753, Jun. 2022, doi: 10.1109/TII.2021.3112386.

[21] C. Jiang, L. Yang, Y. Gao, J. Zhao, W. Hou, and F. Xu, "An Intelligent 5G Unmanned Aerial Vehicle Path Optimization Algorithm for Offshore Wind Farm Inspection," *Drones*, vol. 9, no. 1, 2025, doi: 10.3390/drones9010047.

[22] R. K. Saha and J. M. Cioffi, "Dynamic Spectrum Sharing for 5G NR and 4G LTE Coexistence - A Comprehensive Review," *IEEE Open J. Commun. Soc.*, vol. 5, pp. 795–835, 2024, doi: 10.1109/OJCOMS.2024.3351528.

[23] H. Fehmi, M. F. Amr, A. Bahnasse, and M. Talea, "5G Network: Analysis and Compare 5G NSA /5G SA," *Procedia Comput. Sci.*, vol. 203, pp. 594–598, 2022, doi: 10.1016/j.procs.2022.07.085.

[24] R. Mohamed, S. Zemouri, and C. Verikoukis, "Performance Evaluation and Comparison between SA and NSA 5G Networks in Indoor Environment," in *2021 IEEE International Mediterranean Conference on Communications and Networking (MeditCom)*, Athens, Greece: IEEE, Sep. 2021, pp. 112–116. doi: 10.1109/MeditCom49071.2021.9647621.

[25] "Unified Architecture - Landingpage," OPC Foundation. Accessed: Apr. 15, 2025. [Online]. Available: https://opcfoundation.org/about/opc-technologies/opc-ua/

[26] O. Khalid *et al.*, "Applications of robotics in floating offshore wind farm operations and maintenance: Literature review and trends," *Wind Energy*, vol. 25, no. 11, pp. 1880–1899, 2022, doi: 10.1002/we.2773.

[27] C. Gehring *et al.*, "ANYmal in the Field: Solving Industrial Inspection of an Offshore HVDC Platform with a Quadrupedal Robot," in *Field and Service Robotics*, G. Ishigami and K. Yoshida, Eds., Singapore: Springer, 2021, pp. 247–260. doi: 10.1007/978-981-15-9460-1_18.

[28] B. S. Khan, S. Jangsher, A. Ahmed, and A. Al-Dweik, "URLLC and eMBB in 5G Industrial IoT: A Survey," *IEEE Open J. Commun. Soc.*, vol. 3, pp. 1134–1163, 2022, doi: 10.1109/OJCOMS.2022.3189013.

[29] A. Mwangi, N. Kabbara, P. Coudray, M. Gryning, and M. Gibescu, "Investigating the Dependability of Software-Defined IIoT-Edge Networks for Next-Generation Offshore Wind Farms," *IEEE Trans. Netw. Serv. Manag.*, vol. 21, no. 6, pp. 6126–6139, Dec. 2024, doi: 10.1109/TNSM.2024.3458447.

[30] "Go1 Datasheet_EN v3.0.pdf." Accessed: Mar. 23, 2025. [Online]. Available: https://www.generationrobots.com/media/unitree/Go1%20Datasheet_EN%20v3.0.pdf

[31] M. A. V. Torres and F. Pfitzner, "Investigating Robot Dogs for Construction Monitoring: A Comparative Analysis of Specifications and On-site Requirements," 2023, p. 321 KB, 8 pages. doi: 10.13154/294-10094.

[32] "Go1 — Unitree_Docs 1.0rc documentation." Accessed: Mar. 23, 2025. [Online]. Available: https://unitree-docs.readthedocs.io/en/latest/get_started/Go1_Edu.html#precautions-for-go1-development-using-sdk

[33] "5G RM520N series." Accessed: Apr. 15, 2025. [Online]. Available: https://www.quectel.com/product/5g-rm520n-series/#summary

[34] V. Mannoni, V. Berg, S. Sesia, and E. Perraud, "A Comparison of the V2X Communication Systems: ITS-G5 and C-V2X," in *2019 IEEE 89th Vehicular Technology Conference (VTC2019-Spring)*, Apr. 2019, pp. 1–5. doi: 10.1109/VTCSpring.2019.8746562.

[35] D. Harris-Birtill and R. Harris-Birtill, "Understanding Computation Time: A Critical Discussion of Time as a Computational Performance Metric," in *Time in Variance*, A. Misztal, P. A. Harris, and J. A. Parker, Eds., BRILL, 2021, pp. 220–248. doi: 10.1163/9789004470170_014.

[36] M. Pundir and J. K. Sandhu, "A Systematic Review of Quality of Service in Wireless Sensor Networks using Machine Learning: Recent Trend and Future Vision," *J. Netw. Comput. Appl.*, vol. 188, p. 103084, Aug. 2021, doi: 10.1016/j.jnca.2021.103084.

[37] F. Dettinger, M. Weiß, D. Dittler, J. Stümpfle, M. Artelt, and M. Weyrich, "A Survey on Performance, Current and Future Usage of Vehicle-To-Everything Communication Standards," Oct. 14, 2024, *arXiv*: arXiv:2410.10264. doi: 10.48550/arXiv.2410.10264.